\documentclass[twoside,fleqn]{ActaStyle}
\usepackage{times,cite}

\usepackage{lscape}             % if you need landscape enviroinment
\usepackage{graphicx,epsfig}    % if you need to include figures in PostScript
\usepackage[tbtags]{amsmath}    % if you need more math

\def\authorheading{J.~Smyrski et al.}

\def\shorttitle{Study of the ${^3\mbox{He}}-\eta$ system ...}

\begin{document}
%% ---------------------------------------------------

%%%  page range, first and last page
\pagerange{1}{6}

%%% paper title
\title{%
STUDY OF THE ${^3\mbox{He}}-\eta$ SYSTEM\\
IN $d-p$ COLLISIONS AT COSY-11
}

%%% author(s) and address(es)
\author{%  author(s)
J.~Smyrski\email{smyrski@if.uj.edu.pl}$^a$, H.-H.~Adam\,$^b$,
A.~Budzanowski\,$^c$, E.~Czerwi\'nski\,$^a$, R.~Czy\.zykiewicz\,$^{a,d}$, D.~Gil\,$^a$, D.~Grzonka\,$^d$,
A.~Heczko\,$^a$, M.~Janusz\,$^a$, L.~Jarczyk\,$^a$, B.~Kamys\,$^a$, A.~Khoukaz\,$^b$, K.~Kilian\,$^d$,
P.~Klaja\,$^a$, J.~Majewski\,$^{a,d}$,
P.~Moskal\,$^{a,d}$, W.~Oelert\,$^d$, C.~Piskor-Ignatowicz\,$^a$,
J.~Przerwa\,$^a$, J.~Ritman\,$^d$, T.~Ro\.zek\,$^{d,e}$, R.~Santo\,$^b$, T.~Sefzick\,$^d$,
M.~Siemaszko\,$^e$, A.~T\"aschner\,$^b$, P.~Winter\,$^d$, M.~Wolke\,$^d$,
P.~W\"ustner$^f$, Z.~Zhang\,$^d$, W.~Zipper$^e$
}
{%  address(es)
$^a$M.~Smoluchowski Institute of Physics, Jagellonian University, 30-059 Cracow, Poland \\
$^b$Institut f\"ur Kernphysik, Westf\"alische Wilhelms-Universit\"at, D-48149 M\"unster, Germany\\
$^c$Institute of Nuclear Physics, Pl-31-342 Cracow, Poland\\
$^d$IKP, Forschungszentrum J\"ulich, D-52425 J\"ulich, Germany\\
$^e$Institute of Physics, University of Silesia, PL-40-007 Katowice, Poland\\
$^f$ZEL, Forschungszentrum J\"ulich, D-52425 J\"ulich, Germany
}

%%% Date of submition
\day{November 1, 2005}

%%% abstract of the paper
\abstract{%
We present preliminary results from $dp \rightarrow {^3\mbox{He}}\,X,\, (X=\pi^0, \eta$) measurements 
near the $\eta$ production threshold.
The data were taken during a slow ramping of the COSY internal deuteron beam
scattered on a proton target.
The  ${^3\mbox{He}}$ ejectiles were registered with the COSY-11 detection setup.
The ongoing data analysis should deliver high precision data for the $dp \rightarrow {^3\mbox{He}}\,\eta$ 
total and differential cross sections for the excess energies in the range from threshold up to 9~MeV.
The preliminary excitation function for the reaction $dp \rightarrow {^3\mbox{He}}\,\pi^0$
does not show any structure which could originate from the decay of ${^3\mbox{He}}-\eta$ bound state.
We present also a threshold excitation curve for the $dp \rightarrow {^3\mbox{He}}\,X$ channel.
Contrary to corresponding results from SATURNE we see no cusp in the vicinity of the $\eta$ threshold.
}

%%% PACS numbers of your article
\pacs{%
13.75.Cs, 14.40.-n, 25.45.-z
}

% %%%%%%%%%%%%%%%%%%%%%%%%%%%%%%%%%%%%%%%%%%%%%%%%%%%%%%%%%%%%%%%%%
\section{Introduction}
\label{sec:intr} \setcounter{section}{1}\setcounter{equation}{0}
% %%%%%%%%%%%%%%%%%%%%%%%%%%%%%%%%%%%%%%%%%%%%%%%%%%%%%%%%%%%%%%%%%

One of the basic questions of the $\eta$ meson physics concerns existence
of $\eta$-nucleus bound states postulated by Haider and Liu \cite{1}.
Observation of such states would be very interesting for its own 
but also for studies of $\eta\, N$ interaction and for investigation
of $N^*(1535)$ properties in nuclear matter.
In particular, it could shed a new light on the possible restoration
of chiral symmetry in nuclear matter at normal densities \cite{2,3}.
Liu and Haider \cite{4} claimed that binding of the $\eta$ and a nucleus
should be possible for $A > 10$. 
Later works indicate that this could happen even in the ${^3\mbox{He}} - \eta$ 
system \cite{5,6,7}.
Study of the ${^3\mbox{He}} - \eta$ bound state is of special interest since
it can be described using the Faddeev-Yakubovsky theory \cite{8}.
Recent data from MAMI show some indications for photoproduction of $\eta$-mesic
${^3\mbox{He}}$ and its decay to the $\pi^0 p X$ channel \cite{9}.
However, the data do not allow for an unambiguous   conclusion  whether the observed
enhancement is due to the virtual or bound state~\cite{hanhart}.

For studies of the ${^3\mbox{He}} - \eta$ interaction the $d+p$ collisions
are very well suited due to the relatively high cross sections for the $dp \rightarrow {^3\mbox{He}}\, \eta$
reaction and due to much better beam momentum definition compared 
to the $\gamma {^3\mbox{He}} \rightarrow {^3\mbox{He}}\, \eta$ 
and $\pi^+\,{^3\mbox{H}} \rightarrow {^3\mbox{He}}\, \eta$ measurements.
Of interest are experiments both above and below the $\eta$ production threshold.
In the first case low energy ${^3\mbox{He}} - \eta$ scattering parameters can be determined
on the basis of the FSI effects \cite{10}.
Recently Sibirtsev et al. \cite{11} revised our knowledge of the ${^3\mbox{He}} - \eta$ scattering length
via a systematic study of the available experimental data on the $dp \rightarrow {^3\mbox{He}}\, \eta$
reaction \cite{cosy,wasa,spes_berger,spes_mayer,gem}. 
The authors pointed out several discrepancies between various experiments.
They suggest to perform measurements of angular distributions
at the excess energy of around $Q=6$~MeV in order to examine if the reaction is still dominated
by the s-wave.
They also point out usefulness of measurements very close to threshold for putting
a more stringent constrains on the imaginary part of the scattering length.
In the present experiment we measured the  $dp \rightarrow {^3\mbox{He}}\, \eta$ reaction
from threshold up to $Q=9$~MeV. The experiment and partial results of the ongoing
data analysis are presented in the next chapter.

In measurements below threshold one can search for resonance like structures in excitation curves
originating from decays of ${^3\mbox{He}} - \eta$ bound state 
in various possible reaction channels 
like $dp \rightarrow {^3\mbox{He}}\,\pi^0,{^3\mbox{H}} \pi^+, dp, dp \pi^0, ppp\pi^-$.
In our measurements we registered these channels and preliminary results
for the $dp \rightarrow {^3\mbox{He}}\, \pi^0$ reaction are presented in chapter~3.

An interesting issue which we also studied in the present experiment is the cusp
effect in the threshold excitation curve for the $dp \rightarrow {^3\mbox{He}} X$ process
observed at SATURNE \cite{12}. 
Corresponding results are presented in chapter~4.

% %%%%%%%%%%%%%%%%%%%%%%%%%%%%%%%%%%%%%%%%%%%%%%%%%%%%%%%%%%%%%%%%%
\section{The experimental method}
\label{sec:expe}
% %%%%%%%%%%%%%%%%%%%%%%%%%%%%%%%%%%%%%%%%%%%%%%%%%%%%%%%%%%%%%%%%%

The experiment was performed with the internal deuteron beam of the COSY-J\"ulich accelerator \cite{prashun} scattered
on a proton target of the cluster jet type \cite{dombrowski} and the COSY-11 facility \cite{brauksiepe,jureknim} 
detecting the charged reaction
products.
The choice of the deuteron beam was dictated by the fact that 
the COSY-11 detection acceptance 
for measurements with deuteron beam scattered on a proton target
($dp \rightarrow {^3\mbox{He}}\, \eta$ experiment) is by about a factor of four higher compared
to the acceptance for measurement with proton beam scattered on a deuteron target
($pd \rightarrow {^3\mbox{He}}\, \eta$ experiment).
The momentum of the deuteron beam was varied continuously within each cycle from 3.095~GeV/c
to 3.180~GeV/c, crossing the threshold for the $dp \rightarrow {^3\mbox{He}}\, \eta$ 
reaction at 3.141~GeV/c.
The corresponding variation of the excess energies for the $dp \rightarrow {^3\mbox{He}}\, \eta$ reaction
lies in the range from -10~MeV to 9~MeV.
The accuracy of absolute beam momenta was about 3~MeV/c 
and the corresponding uncertainty of the excess energy is 0.7~MeV.
The momentum profile of the deuteron beam determined on the basis of the monitored frequency
spectrum of the COSY accelerator had a total width of 1.7~MeV/c. The corresponding smearing 
of the excess energy $Q$ for the $dp \rightarrow {^3\mbox{He}} \eta$ reaction is equal to 0.4~MeV.
We expect that it should be possible to determine the lowest $dp \rightarrow {^3\mbox{He}} \eta$ data point 
at an excess energy equal roughly to half of this value ($Q = 0.2$~MeV).

The data taking during the ramping phase of the beam was already successfully conducted using the COSY-11 facility \cite{smyrski,moskal_1998}.
Application of this technique allows for the reduction 
of most of the systematic errors associated with relative normalization of points
measured at different beam momenta and appearing in the case when the beam is set up
for each point separately. 
As the most serious source of systematic errors we consider displacement of the beam position
at the target correlated with variation of the beam momentum. 
Therefore, we monitored the beam position in the horizontal direction
using measurements of $p-p$ quasi elastic scattering 
and applying methods described in reference \cite{moskal_2001}. In the vertical plane we used the reconstruction of 
the reaction vertices by tracing particle trajectories
in the magnetic field of the COSY-11 dipole magnet.

\begin{figure}[htb]
\begin{center}
\includegraphics[width=2.5in]{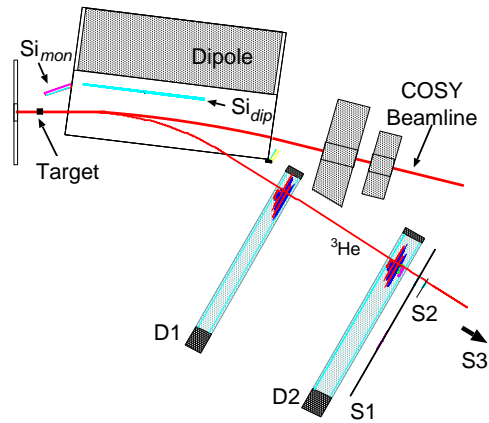} \\
\end{center}
\caption{%
Schematic view of the COSY-11 detection system \cite{brauksiepe}.
}
\label{fig:1}
\end{figure}

A schematic view of the experiment is shown in Fig.~\ref{fig:1}.
The ${^3\mbox{He}}$ ejectiles were momentum analyzed in the COSY-11 dipole magnet and their trajectories 
were registered in  two drift chambers D1 and D2. 
For particle identification, the time of flight on a path of 9~m 
between the start scintillation hodoscope S1 and the stop hodoscope S3 was measured.
%  Fig. 1

\begin{figure}[htb]
\begin{center}
\includegraphics[width=5.0in]{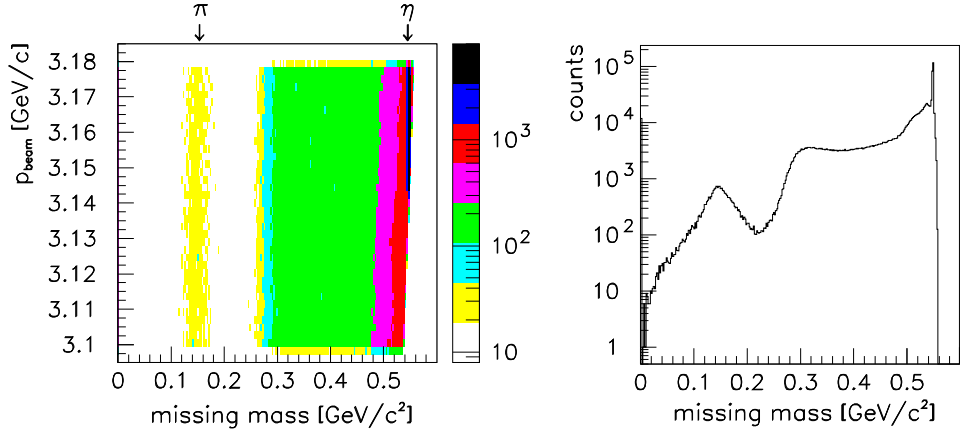} \\
\end{center}
\caption{
Left plot: missing mass (x-axis) as a function of beam momentum (y-axis).
Right plot: missing mass distribution obtained as a projection of the left plot
on the missing mass axis.
}
\label{fig:2}
\end{figure}

In the missing mass spectrum determined as a function of the beam momentum (see Fig.~\ref{fig:2}) a clear
signals from the $\eta$ meson production as well as from the single $\pi^0$ production are visible.
The background under the $\eta$ peak can be very well reproduced and subtracted
on the basis of measurements below threshold prescaled according to the monitored luminosity
and the kinematical limit for the missing mass.
The ongoing data analysis indicates that the collected data will allow to determine the total and differential
cross sections for the $dp \rightarrow {^3\mbox{He}} \eta$ reaction with a high statistical accuracy
on the level of 1\%.

%  Fig. 2

\begin{figure}[htb]
\begin{center}
\includegraphics[width=2.5in]{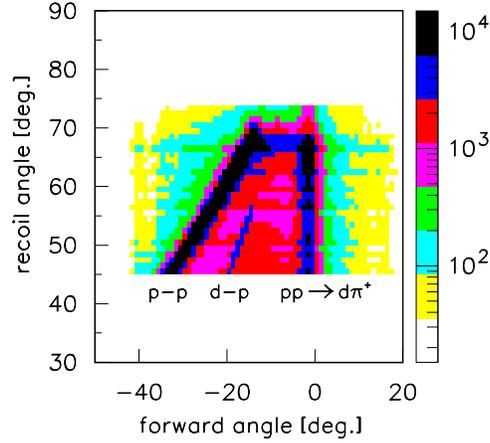} \\
\end{center}
\caption{%
Correlation between the forward and recoil scattering angle.
Clear signals from proton-proton quasi-elastic scattering, 
deuteron-proton elastic scattering
and $pp \rightarrow d \pi^+$ quasi-free process are visible.
}
\label{fig:3}
\end{figure}

During the experiment, the luminosity was monitored using coincident measurement of the elastic $d-p$ scattering
as well as of the $p-p$ and $pp \rightarrow d \pi^+$ quasi-free scattering with the forward
scattered particles registered in the drift chambers and the recoil particles measured
with the silicon strip detectors. The corresponding correlations observed between the forward and recoil angles
are shown in Fig.~\ref{fig:3}.
%  Fig. 3

% %%%%%%%%%%%%%%%%%%%%%%%%%%%%%%%%%%%%%%%%%%%%%%%%%%%%%%%%%%%%%%%%%
\section{Pion production at the eta threshold}
\label{sec:pion}
% %%%%%%%%%%%%%%%%%%%%%%%%%%%%%%%%%%%%%%%%%%%%%%%%%%%%%%%%%%%%%%%%%

For a search of a resonance-like structure originating from decays of ${^3\mbox{He}}-\eta$ 
bound state in the $dp \rightarrow {^3\mbox{He}}\, \pi^0$ channel
we concentrate on angular range of pions covering the forward angles in the $dp$
center-of-mass (CM) system ($\Theta_{d-\pi}^{CM} <90^{\circ}$). 
This choice is dictated by the fact that the $dp \rightarrow {^3\mbox{He}} \pi^0$
cross section is up to two orders of magnitude smaller in this range than at the backward angles \cite{13}.
One can expect that due to the multiple rescattering of the $\eta$ meson
off nucleons, the $\eta$-nucleus "forgets" the orientation of the beam momentum
and, therefore, it decays isotropically.
For this reason it can be best seen  just in the above angular range since
it appears on the level of small ``non-resonant'' cross section.

\begin{figure}[htb]
\begin{center}
\includegraphics[width=2.5in]{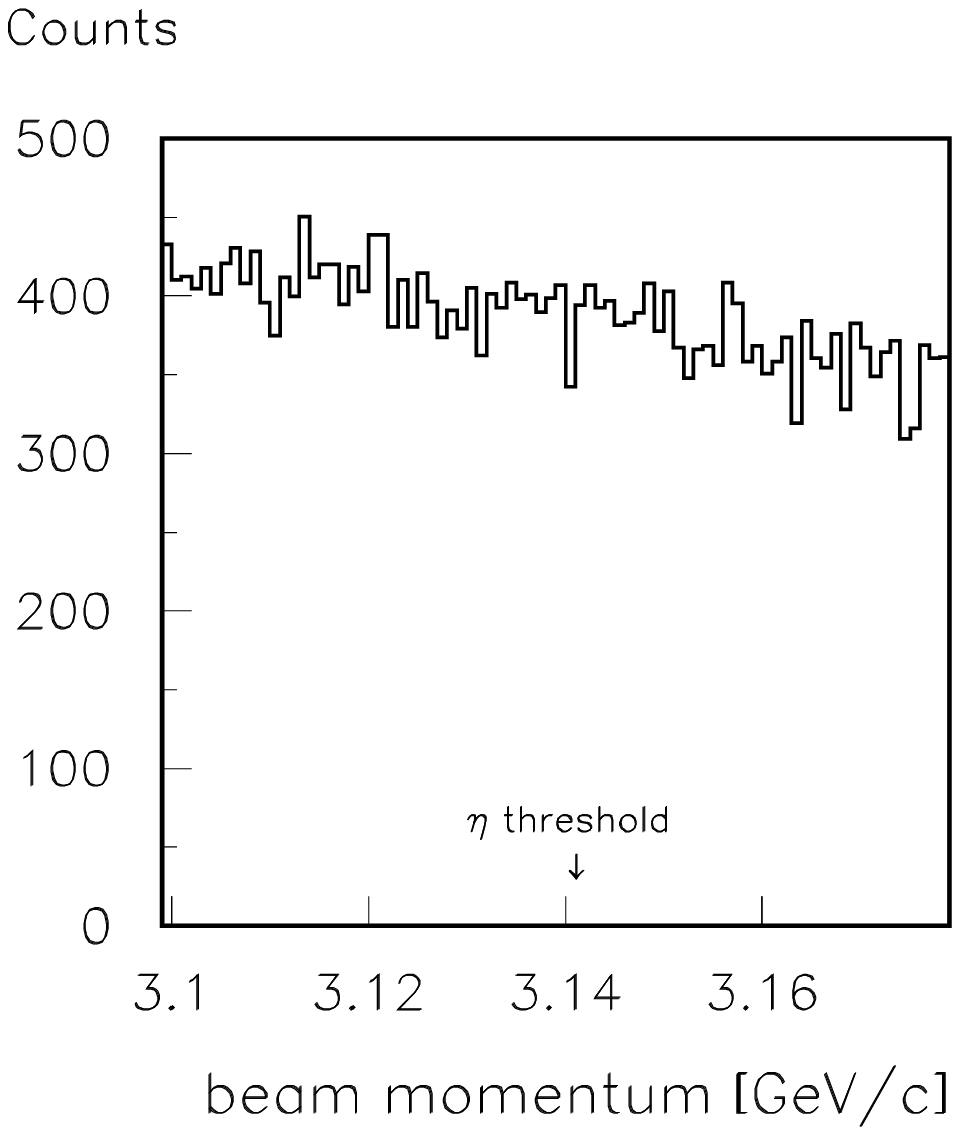} \\
\end{center}
\caption{%
Counting rate %pions from 
of the $dp \rightarrow {^3\mbox{He}} \pi^0$ production
at the forward angles ($\Theta_{d-\pi}^{CM} <90^{\circ}$) as a function of the beam momentum.
}
\label{fig:4}
\end{figure}

Fig.~\ref{fig:4} shows pion counts as a function of the beam momentum determined
after analyzing about 60\% of the collected data.
%  Fig. 4
Except of statistical fluctuations no structure can be seen in this curve.
This can not be treated, however, as a proof of non existence of the ${^3\mbox{He}}-\eta$
bound state since simple estimations lead to the conclusion that the signal from decays of such state
can be too week to be seen in the present experiment. 
Assuming namely, that the cross sections for the ${^3\mbox{He}}-\eta$ bound state formation are of the same
order as the $dp \rightarrow {^3\mbox{He}}\, \eta$ cross sections near threshold (0.4$\mu$b), the decay
probability of the ${^3\mbox{He}}-\eta$ bound state in the ${^3\mbox{He}} \pi^0$ channel is 0.01 
and taking into account that the acceptance of COSY-11 facility for measuring $dp \rightarrow {^3\mbox{He}} \pi^0$ channel
at the forward angles is equal to 0.05, then one expects a signal of only 5 counts per beam momentum bin of 1~MeV/c 
appearing on the observed level of 400 counts originating from non-resonant background (see Fig.~\ref{fig:4}).
We assumed relatively small decay probability in the ${^3\mbox{He}}-\pi^0$ channel (0.01) 
since one can expect that 
the decays of the ${^3\mbox{He}}-\eta$ bound state will be dominated by two body processes
of conversion of $\eta$ mesons on a proton or a neutron inside the  ${^3\mbox{He}}$ nucleus
leading to production of nucleon and pion pair emitted back-to-back.
Therefore, one can expect, that $dp \rightarrow dp\,\pi^0$ process should be more sensitive for studying existence
of ${^3\mbox{He}}-\eta$ bound state. Analysis of this channel is in progress.

% %%%%%%%%%%%%%%%%%%%%%%%%%%%%%%%%%%%%%%%%%%%%%%%%%%%%%%%%%%%%%%%%%
\section{Threshold excitation curve}
\label{sec:exci}
% %%%%%%%%%%%%%%%%%%%%%%%%%%%%%%%%%%%%%%%%%%%%%%%%%%%%%%%%%%%%%%%%%

\begin{figure}[htb]
\begin{center}
\includegraphics[width=2.5in]{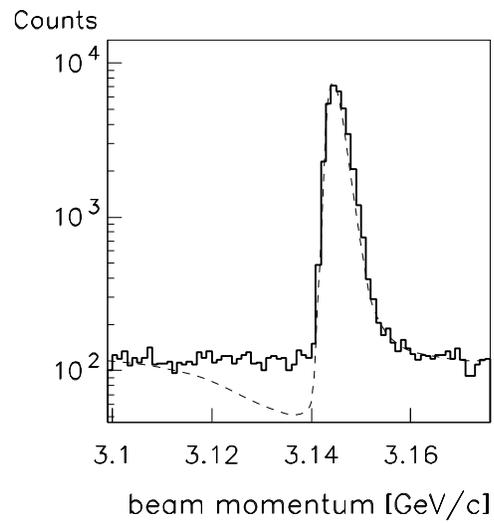} \\
\end{center}
\caption{%
Threshold excitation curve for the $dp \rightarrow {^3\mbox{He}}\, X$ reaction
measured in the neighborhood of the $\eta$ threshold (solid line) and expectation
of the shape of this curve with the cusp structure expected from Ref.~\cite{12} (dashed line). 
}
\label{fig:5}
\end{figure}

Data collected in the present experiment 
has been also used to investigate the cusp effect observed at SATURNE in the 
threshold excitation curve for the process $dp \rightarrow {^3\mbox{He}}\, X$ \cite{12}.
This cusp was visible at the $\eta$ threshold and to the best of our knowledge
there were no other experimental attempts to confirm it.
As suggested by Wilkin \cite{14} it can be caused by an interference between an intermediate state including
the $\eta$ meson and the non-resonant background corresponding to the multi-pion production.
In the SATURNE experiment the ${^3\mbox{He}}$ ejectiles were measured with the SPES~IV spectrometer
which was set in such a way that it registered the ${^3\mbox{He}}$ which was approximately at rest 
in the CM frame. 
The longitudinal (transverse) CM momentum spread of the ${^3\mbox{He}}$  
was equal to $\pm$35~MeV/c ($\pm$10~MeV/c) and was fixed by the momentum (angular) acceptance
of the spectrometer.
The corresponding acceptance in the missing mass $m_X$ was $\Delta m_X \approx$ 1~MeV/c$^2$ and it covered
the range of the highest produced masses at a given beam momentum.
The threshold excitation curve was determined by scanning different masses $m_X$ by varying
the beam momentum and adjusting the setting of the spectrometer in such a way that only the ${^3\mbox{He}}$
associated with the maximum missing masses were registered. 
Since the COSY-11 momentum and angular acceptance is much larger than one of the SPES~IV spectrometer,
the limitation on the c.m. momenta of ${^3\mbox{He}}$ was realized by means of corresponding cuts 
during the data analysis.
For this we determined counts in the missing mass range ($m_{max}-\Delta m, m_{max}$)
where $m_{max}$ is the maximum kinematically allowed missing mass at a given beam momentum
and $\Delta m$ is the width of a bin in the missing mass used for the scan. In order to reproduce the conditions of the SPES~IV we took $\Delta m =$ 1~MeV/c$^2$ .
These counts are shown in Fig.~\ref{fig:5} with a solid line as a function of the beam momentum.
The peak in the middle of the spectrum is associated with opening 
of the $dp \rightarrow {^3\mbox{He}}\, \eta$ channel. 
The dashed line in Fig.\ref{fig:5} represents the expected threshold excitation
curve showing qualitatively the shape of the cusp structure which we were looking for.
Contrary to the SATURNE result with the cusp amplitude reaching about 50\%
of the non-resonant background we see no cusp near the $\eta$ threshold.

%  Fig. 5

% %%%%%%%%%%%%%%%%%%%%%%%%%%%%%%%%%%%%%%%%%%%%%%%%%%%%%%%%%%%%%%%%%
\section{Summary and outlook}
\label{sec:conc}
% %%%%%%%%%%%%%%%%%%%%%%%%%%%%%%%%%%%%%%%%%%%%%%%%%%%%%%%%%%%%%%%%%

We performed measurements of the $dp \rightarrow {^3\mbox{He}}\,X,\, X=\pi^0, \eta$ reactions
near the $\eta$ threshold using a slowly ramped deuteron beam of the COSY-J\"ulich synchrotron.
Determination of the total and differential cross sections for the $dp \rightarrow {^3\mbox{He}}\,\eta$ channel
for the excess energies in the range from threshold up to $Q=9.0$~MeV is in progress.
The excitation curve for pion production in the reaction $dp \rightarrow {^3\mbox{He}}\,\pi^0$
shows no structure originating from decays of a possible ${^3\mbox{He}}-\eta$ bound state.
This can be explained e.g. by a small decay probability of ${^3\mbox{He}}-\eta$ bound state
in the ${^3\mbox{He}}\,\pi^0$ channel.
We measured also the threshold excitation curve for the $dp \rightarrow {^3\mbox{He}}\,X$ process,
however, contrary to the SATURNE results, we observe no cusp near the $\eta$ threshold.
We consider to extend the present experiment on the reactions induced by a deuteron beam on a deuteron target.
The primary goal of these measurements is a search for the ${^4\mbox{He}}-\eta$ bound state via
a measurement of the excitation function for the $dd \rightarrow {^3\mbox{He}}p\pi^-$ reaction
where the outgoing $p-\pi^-$ pair originates from the conversion of the $\eta$ meson on a neutron
inside the ${^4\mbox{He}}$ nucleus and the ${^3\mbox{He}}$ ejectile is an ``observer''.

\begin{ack}
We acknowledge the support of the
European Community-Research Infrastructure Activity
under the FP6 "Structuring the European Research Area" programme
(HadronPhysics, contract number RII3-CT-2004-506078),
of the FFE grants (41266606 and 41266654) from the Research Centre J{\"u}lich,
of the DAAD Exchange Programme (PPP-Polen),
of the Polish State Committe for Scientific Research
(grant No. PB1060/P03/2004/26), \\
and of the
RII3/CT/2004/506078 - Hadron Physics-Activity -N4:EtaMesonNet.
\end{ack}

\end{document}